\documentclass[11pt]{article}
\newcommand{\Comment}[1]{{}}
\usepackage{amsfonts,amsthm,amsmath,amssymb,slashed}
\usepackage[textwidth = 430 pt, textheight = 630 pt]{geometry}
\usepackage{color}

\Comment{\usepackage{color}
\definecolor{MyDarkBlue}{rgb}{0.15,0.15,0.45}
\usepackage[linktocpage=true]{hyperref}
\hypersetup{
colorlinks=true,
citecolor=MyDarkBlue,
linkcolor=MyDarkBlue,
urlcolor=MyDarkBlue,
pdfauthor={Carlos Cardona, Cristhiam Lopez-Arcos and Horatiu Nastase},
pdftitle={Abelian reductions of ${\cal N}=4$ SYM},
pdfsubject={hep-th}
}

\usepackage[numbers,sort&compress]{natbib}
\usepackage{hypernat}}
\usepackage{graphicx}
\usepackage{cite}

\newcommand\ignore[1]{}
\def\one{{\,\hbox{1\kern-.8mm l}}}

\def\Tr{{\rm Tr\, }}

 \newcommand{\pd}{\partial}

\def\a{\alpha}

\def\pb{\ov\psi}
\def\pb{|\phi|}\def\vp{\varphi}
\def\d{\partial}

\def\Tr{\mathop{\rm Tr}\nolimits}

\newcommand{\Cset}{{\,\,{{{^{_{\pmb{\mid}}}}\kern-.45em{\mathrm C}}}}}

\newcommand{\be}{\begin{equation}}
\newcommand{\bea}{\begin{eqnarray}}

\newcommand{\ee}{\end{equation}}
\newcommand{\eea}{\end{eqnarray}}

\newcommand{\nn}{\nonumber}

\begin{document}

\renewcommand{\thefootnote}{\fnsymbol{footnote}}

\makeatletter
\@addtoreset{equation}{section}
\makeatother
\renewcommand{\theequation}{\thesection.\arabic{equation}}

\rightline{}
\rightline{}

\begin{flushright}
QGASLAB-15-01
\end{flushright}

\vspace{10pt}


\begin{center}
{\LARGE \bf{\sc Abelian reductions of deformed ${\cal N}=4$ SYM}}
\end{center} 
 \vspace{1truecm}
\thispagestyle{empty} \centerline{
{\large \bf {\sc Carlos Cardona${}^{a,}$}}\footnote{E-mail address: \Comment{\href{mailto:cargicar@ift.unesp.br}}{\tt cargicar@ift.unesp.br}},
{\large \bf {\sc Cristhiam Lopez-Arcos${}^{b,}$}}\footnote{E-mail address: \Comment{\href{mailto:crismalo@ift.unesp.br}}{\tt crismalo@ift.unesp.br}}
{\bf{\sc and}}
{\large \bf {\sc Horatiu Nastase${}^{a,}$}}\footnote{E-mail address: \Comment{\href{mailto:nastase@ift.unesp.br}}{\tt nastase@ift.unesp.br}}
                                                           }

\vspace{.5cm}


\centerline{{\it ${}^a$ 
Instituto de F\'{i}sica Te\'{o}rica, UNESP-Universidade Estadual Paulista}} \centerline{{\it 
R. Dr. Bento T. Ferraz 271, Bl. II, Sao Paulo 01140-070, SP, Brazil}}

\centerline{{\it ${}^b$
The Laboratory for Quantum Gravity \& Strings, }} \centerline{{\it
Department of Mathematics and Applied Mathematics, }} \centerline{{\it
University of Cape Town, Private Bag, Rondebosch, 7700, South Africa}}

\vspace{1truecm}

\thispagestyle{empty}

\centerline{\sc Abstract}

\vspace{.4truecm}

\begin{center}
\begin{minipage}[c]{380pt}
{\noindent Following the work in \cite{Mohammed:2012gi}, where the massive ABJM model in 2+1 dimensions was shown to have an abelian reduction 
to the relativistic Landau-Ginzburg, and motivated by the implications for condensed matter through AdS/CFT, 
we show that a FI deformation of ${\cal N}=4$ SYM in 3+1 dimensions with a mass term can also be reduced to a
relativistic Landau-Ginzburg model, with the possibility of coupling it to a real scalar, whereas the simply mass deformed ${\cal N}=4$ SYM reduces only 
to a massive $\phi^4$ model (scalar QED) coupled to a real scalar. We study the classical solutions of the model, in particular vortex solutions.
}
\end{minipage}
\end{center}

\vspace{.5cm}

\setcounter{page}{0}
\setcounter{tocdepth}{2}

\newpage

\renewcommand{\thefootnote}{\arabic{footnote}}
\setcounter{footnote}{0}

\linespread{1.1}
\parskip 4pt



\section{Introduction}
\ \ \ \ \ Over the last decade, applications of AdS/CFT \cite{Maldacena:1997re}
 to condensed matter ("AdS/CMT") have become very popular (see, e.g., the review \cite{Hartnoll:2009sz}). 
 Usually one takes a "bottom-up" point of 
view, and constructs a gravitational theory in AdS space which has desirable features for some field theory dual operators, without knowing what 
the field theory is, and if it is really related to the condensed matter system of interest. Moreover, usually the condensed matter system is described 
by some abelian effective field theory, unlike the field theory dual to AdS space, which needs a large number of fields, organized in some large $N$ 
matrix. Another approach is a "top-down" one, in which some gravity dual pair coming from string theory is applied to some condensed matter problem, 
but usually just as a toy model. Moreover, the same issue applies, with the effective description of the condensed matter system being usually abelian.

In \cite{Mohammed:2012gi,Mohammed:2012rd}, a step was taken towards a better foundation for AdS/CMT applications, by taking a bit of both 
approaches. One takes a top-down model, and sees whether it has an abelian reduction that is an effective field theory model for condensed 
matter. In the case in \cite{Mohammed:2012gi,Mohammed:2012rd}, the massively deformed \cite{Gomis:2008vc} (see also \cite{Terashima:2008sy})
2+1 dimensional ABJM model \cite{Aharony:2008ug} was shown to reduce
in a nontrivial way, that still preserves the gravity dual,  to a relativistic 2+1 dimensional
Landau-Ginzburg model that was for instance used to describe the quantum critical phase \cite{Myers:2010pk,Sachdev:2011wg}. 
Moreover, the reduction was shown to be 
a consistent truncation, that can even be made consistent at the quantum level, provided one takes a fine-tuned region of parameter space, 
and the reduction simulates the reduction in degrees of freedom happening in a condensed matter system when one derives the LG effective
field theory. 

Many condensed matter systems of interest, in particular for AdS/CFT, effectively live in 2+1 dimensions, but in any case, the general system is 
always 3+1 dimensional. It is therefore of interest to see if a similar story applies in 3+1 dimensions.

In this paper, we study possible deformations of ${\cal N}=4$ SYM, the standard toy model in 3+1 dimensions, for which AdS/CFT is best 
understood, and check whether there is a possible reduction to the relativistic Landau-Ginzburg model. We find that a simple mass deformation 
does not allow the possibility of reduction to LG, but if we add also a FI deformation it does. Moreover, one can have also a coupling to a nontrivial real scalar.
However, we analyze vortex solution ans\"{a}tze, and prove that there are no vortex solutions other than the usual Abrikosov-Nielsen-Olesen \cite{Abrikosov:1956sx,Nielsen:1973cs} ones, 
be they BPS or non-BPS. 

The paper is organized as follows. In section 2 we analyze mass deformations and truncation ans\"{a}tze, and see that they don't lead to 
the LG model, but simply to a massive $\phi^4$ scalar (there is no possibility of symmetry breaking), coupled to a real scalar. 
In section 3 we analyze FI deformations and 
and show that in this case we can reduce to the LG model, plus a coupling to a real scalar. In section 4 we analyze the LG plus real scalar model, 
and show that the only vortices, in either BPS or non-BPS cases, are the ones with the real scalar put to zero, i.e. the usual Abrikosov-Nielsen-Olesen vortices, 
and in section 5 we conclude.

\section{Mass deformations of ${\cal N}=4$ SYM and truncation ans\"{a}tze. Reduction to scalar QED coupled to real scalar.}

\subsection{Single mass deformation}

{\bf Action and vacuum}

The bosonic Lagrangean for ${\cal N}=4$ SYM in Euclidean space is 
\begin{eqnarray}
{\cal L}_{{\cal N} = 4} &=& \frac{1}{2\,g^2}{\rm Tr} \, \left\{
\frac{1}{2}F_{\mu\nu} F^{\mu\nu} +\left( { D}_\mu X^{I} \right)
\left( {D}^\mu X_{I} \right) - \sum_{I<J=1}^6[X^{I},X^{J}] [X_{I}, X_{J}] \right\}\,.  \label{lagrangian}
\end{eqnarray}
Here we used Hermitian generators, with the normalization $\Tr[T^a T^b]=+\delta^{ab}$ and $D^\mu =\d^\mu -i[A^\mu,.]$.\footnote{Note that 
$\sum_{I<J} [X^I,X^J]^2=1/2\sum_{I,J}[X^I,X^J]^2$ and $\overline{X^I}=X^I$, so $\overline{[X^I,X^J]}=-[X^I,X^J]$.}
The scalars $X^I$ transform in the fundamental of $SO(6)_R$ R-symmetry, and both $X^I$ and the gauge fields $A_\mu$ are in the adjoint 
of $SU(N)$.

We first consider a mass deformation with a single mass parameter, that preserves ${\cal N}=1$ supersymmetry (see for instance \cite{Vafa:1994tf}). 
Forming the complex combinations $\Phi_m=X_m+iX_{m+3}$ for $m=1,2,3$, and promoting them to superfields, the superpotential 
in ${\cal N}=1$ notation is 
\be 
W=-im\Tr(\Phi_I\Phi^I)+\Tr(\Phi_1[\Phi_2,\Phi_3]).
\ee
Note that the mass parameter $m$ is a priori complex, but we chose it to be purely imaginary ($im$), having in mind the particular vacuum we want to 
study. Then the bosonic part of the action with the mass deformation is 
\bea\label{lbos}
{\cal L}_{Bos} &=& \frac{1}{2g^2}{\rm Tr} \bigg\{\frac{1}{2}
F_{\mu\nu} F^{\mu\nu} + \left( { D}_\mu \Phi^{m} \right)\left( {D}^\mu \overline{\Phi}_{m} \right)+\left|\frac{[\Phi_1,\overline{\Phi_1}]}{2}\right|^2
+\left|\frac{[\Phi_2,\overline{\Phi_2}]}{2}\right|^2+\left|\frac{[\Phi_3,\overline{\Phi_3}]}{2}\right|^2
\nn\\
&&+ |[\Phi_2,\Phi_3]-2im\Phi_1|^2+|[\Phi_3,\Phi_1]-2im\Phi_2|^2+|[\Phi_1,\Phi_2]-2im\Phi_3|^2\bigg\}\nn\\
&=& \frac{1}{2g^2}{\rm Tr} \bigg\{\frac{1}{2}
F_{\mu\nu} F^{\mu\nu} + \left({D}_\mu\Phi^{m}\right)\left({D}^\mu\overline{\Phi}_{m}\right) + 4m^2\sum_{m=1}^3|\Phi_m|^2
+\sum_{m=1}^3\left|\frac{[\Phi_m,\overline{\Phi_m}]}{2}\right|^2\nn\\
&&+ 2im\Big(-\Phi_1\overline{[\Phi_2,\Phi_3]} + \overline{\Phi}_1[\Phi_2,\Phi_3] - \Phi_2\overline{[\Phi_3,\Phi_1]} + \overline{\Phi}_2[\Phi_3,\Phi_1]\nn\\ 
&&- \Phi_3\overline{[\Phi_1,\Phi_2]} + \overline{\Phi}_3[\Phi_1,\Phi_2]\Big) + \sum_{m<n=1}^3|[\Phi_m,\Phi_n]|^2\bigg\}.
\eea
In the potential, the $|[\Phi_m,\overline{\Phi}_m]|^2$ terms are D-terms $|D^a|^2=|\Phi T^a \tilde \Phi|^2$, and the rest are F-terms, coming from the 
superpotential.\footnote{Note that as before, $\overline{[\Phi_m,\Phi_n]}=-[\overline{\Phi}_m,\overline{\Phi}_n]$, so 
$|[\Phi_m,\Phi_n]|^2=[\Phi_m,\Phi_n]\overline{[\Phi^m,\Phi^n]}=-[\Phi_m,\Phi_n][\overline{\Phi^m},\overline{\Phi^n}]\geq 0$.}

A supersymmetric vacuum solution has the D-terms equal to zero by taking $X_4=X_5=X_6=0$, i.e. $\Phi_m$ real, and the F-terms to zero, 
giving
\be
[\Phi_m,\Phi_n]=2mi\epsilon_{mnp}\Phi_p\;,
\ee
where now $\Phi_m=X_m$ is real (Hermitean). The solutions to these equations are $N$-dimensional matrix representations of the $SU(2)$ 
Lie algebra, i.e. a fuzzy 2-sphere, with radius $r^2\propto N^2$. 

{\bf Truncation}

Consider therefore the truncation $X_{m+3}=0$, $\Phi_m$ real (Hermitean) of the bosonic Lagrangean, which can be easily checked to be 
consistent (there are no linear terms in $X_{m+3}$ in the action, on the ansatz) leading to 
\bea\label{lbost}
{\cal L}_{Bos} &=& \frac{1}{2g^2}{\rm Tr} \Bigg\{ \frac{1}{2}
F_{\mu\nu} F^{\mu\nu} + \left( { D}_\mu X^{m} \right)
\left( {D}^\mu X_{m} \right)\nn\\
&& + 4m^2\sum_{m=1}^3X_{m}^2 + 12imX_3[X_1,X_2] - \sum_{m<n=1}^3[X_m,X_n]^2\frac{}{}\Bigg\}\;,
\eea
and make the $SU(2)$-inspired combinations $X^\pm=X_1\pm i X_2$, $X_3$ real and rescale $X_m\rightarrow gX_m$, $A_\mu\rightarrow g A_\mu$,
leading to the bosonic
Lagrangean
\bea
{\cal L}_{Bos} &=& {\rm Tr} \Bigg\{ \frac{1}{4}
F_{\mu\nu} F^{\mu\nu} + \frac{1}{2}\left( { D}_\mu X_{3} \right) \left( {D}^\mu X_{3} \right) + \frac{1}{2}\left( { D}_\mu X^{+} \right) \left( {D}^\mu X^{-} \right)\nn\\
&& + 2m^2\left(X_{3}^2 + X^{+}X^{-}\right) + 3mgX_3[X^+,X^-]\nn\\
&& - \frac{g^2}{2}\left(-\frac{1}{4}[X^+,X^-]^2 + [X^+,X_3][X^-,X_3]\right)\Bigg\}.\label{lbospm}
\eea
Defining the usual generators of $SU(2)$, $[J^i,J^j]=i\epsilon^{ijk}J^k$, rewritten as
$J^{\pm} = J^1 \pm iJ^2$ and $J^3$, satisfying
\be
\left[J^3,J^{\pm}\right] = \pm J^{\pm},\qquad
\left[J^{+},J^{-}\right] = 2J^3\;,
\ee    
so $\Tr[J^+J^-]=2N$, where $N$ is the dimension of the $SU(2)$ representation,  
we can easily define an abelianization ansatz 
\be
X^+=\phi J^+;\;\;\;
X^-=\phi^* J^-;\;\;\;
X_3=\chi J_3;\;\;\;\
A_\mu=a_\mu J_3.
\ee
Note that this reduction "embeds part of the R-symmetry in the gauge group", by identifying $m$ indices with an $SU(2)$ subgroup of the 
$a$ indices.  
We have verified that this, together with the variant from section 4, is the only nontrivial 
consistent truncation ansatz in terms of the $SU(2)$ generators $J_i$ that involves a complex scalar and a gauge field.
The covariant derivatives reduce as
\bea
D_\mu X^+&=&\d_\mu \phi J^+-iga_\mu \phi [J_3,J^+]=(\d_\mu \phi -ig a_\mu \phi)J^+,\cr
D_\mu X^-&=&\d_\mu \phi J^--iga_\mu \phi^*[J_3,J^-]=(\d_\mu \phi^*+iga_\mu \phi^*)J^-,\cr
D_\mu X_3&=&\d_\mu \chi J_3-iga_\mu \chi [J_3,J_3]=\d_\mu \chi J_3\;,
\eea
and the potential terms (quartic, cubic and mass, respectively) reduce as
\bea
\frac{g^2}{2}{\rm Tr}\left(-\frac{1}{4}[X^+,X^-]^2 + [X^+,X_3][X^-,X_3]\right)&=&-N\frac{g^2}{2}(|\phi|^4 + 2|\phi|^2\chi^2),\nn\\
3mg{\rm Tr}\left(X_3[X^+,X^-]\right)&=&6Nmg|\phi|^2\chi,\nn\\
2m^2{\rm Tr}\left(X_{3}^2 + X^{+}X^{-}\right)&=& 2Nm^2(2|\phi|^2 +\chi^2)).
\eea
The field strength reduces simply to the abelian one, $F_{\mu\nu}= (\partial_{\mu}A_{\nu} - \partial_{\nu}A_{\mu})J_3$.

Putting all the terms together, we obtain the reduced abelian action
\bea\label{abels}
S&=&N\int d^4x \bigg[+\frac{1}{4}
F_{\mu\nu} F^{\mu\nu} + \left( {D}_\mu \phi \right) \left(\overline{{D}^\mu\phi}\right) + \frac{1}{2}(\partial_{\mu}\chi)^2 + 2m^2(\chi^2 + 2|\phi|^2)\nn\\
&&\hspace{1cm} + 6mg|\phi|^2\chi + \frac{1}{2}g^2(\vert\phi\vert^4 + 2\vert\phi\vert^2\chi^2)\bigg].
\eea
If we would put $\chi=0$, we would obtain simply massive scalar QED, i.e. a gauge field coupled to a massive complex scalar with $\phi^4$ interaction. 
However, note that putting $\chi=0$ (or equal to any other constant) in the above is not a consistent truncation, since we have a linear term in $\chi$ 
in the action, thus a nonzero source term (for nonzero $\phi$) for the $\chi$ equation of motion. 

{\bf Consistency of the truncation}

To verify the consistency of the truncation, we write the equations of motion of the original action (\ref{lbost}), or equivalently (\ref{lbospm}), 
and see if they are satisfied by the equations of motion for the truncation ansatz. 

The equation of motion for $X^-$ is
\be
-\frac{1}{2}D^2 X^+ + 2m^2X^+ + 3mg[X_3,X^+] + g^2\left(\frac{1}{4}[[X^+,X^-],X^+] - \frac{1}{2}[[X_3,X^+],X_3]\right) = 0\;,
\ee
and on our ansatz, it reduces to 
\be
J^+\left[-\frac{1}{2}D^2\phi + 2m^2\phi + 3mg\chi\phi + \frac{1}{2}g^2(|\phi|^2\phi + \chi^2\phi)\right] = 0,\label{phieq}
\ee
which is the reduced equation of motion for $\phi^*$ from the abelian action \eqref{abels} (times a global factor of 1/2). 

Similarly, the equation of motion for $X^+$ reduces to the equation of motion for $\phi$. 

The equation of motion for $X_3$ is 
\be
-D^2X_3 + 4m^2X_3 + 3mg[X^+,X^-] - g^2[[X^-,X_3],X^+] = 0\;,
\ee
and it reduces on our ansatz to 
\be
J_3[-\d^2\chi + 4m^2\chi + 6mg|\phi|^2 + 2g^2|\phi|^2\chi] = 0,\label{chieq}
\ee
which is the equation of motion for $\chi$. 

Finally, the equation of motion for $A^a_\mu$ reduces to 
\be\label{Aeom}
\d^2A^a_\mu -\d^\nu (\d^\mu A^a_\mu)+g\delta^{mn}X_m^b D_{\mu}X_n^cf^a_{~bc}=0\;,
\ee
where we have written explicitly both the R-symmetry $m$ indices and the gauge $a$ indices. The ansatz has only $m=b$, $n=c$ components nonzero. 
For $a=3$, the gauge equation of motion reduces on our ansatz to 
\be
J_3[(\d^2a_\mu -\d_\mu (\d^\nu a_\nu))-ig(\phi (D_\mu \phi)^*-\phi^*D_\mu \phi)]=0\;,\label{gaugeeq}
\ee
which is the equation of motion for $a_\mu$. 

For $a=+$ and $a=-$, we get 0=0.
Indeed, note that keeping 
$A_\mu^+J^++A_\mu^-J^-=2(A_\mu^1J^1+A_\mu^2 J^2)$ nonzero, where $A_\mu^\pm=A_\mu^1\mp iA_\mu^2$,
we would get in the Lagrangean the terms 
\be
\delta^{mn}(D_\mu X_m)^-(\d_\mu X_n^+-iA_\mu^3X_n^+ +i A_\mu^+ X_n^3)+\delta^{mn}(\d_\mu X_m)^3(\d_\mu X_n^3-iA_\mu^+X_n^-+i A_\mu^- X_n^+)\;,
\ee
but the terms with $A^+_\mu$ contain $D_\mu X_m^- X_n^3$ and $D_\mu X_m^3 X_n^-$, which are zero on the ansatz.

\Comment{we obtain the condition
\be
\phi^* \d_\mu \chi-\chi D_\mu \phi^*=0\;,\label{constraint}
\ee
which is not part of the equations of motion of the reduced action, so must be imposed as a constraint.}

{\bf Vacuum and pure scalar solutions}

We see that simply putting $\chi=$ b = constant is not a solution, unless we put $-\chi=|\phi|=m/g$, which is the fuzzy sphere ground state.

Indeed, the vacuum solution (for constant fields) is found from the equations
\bea
&&2m^2\phi + 3mg\chi\phi + \frac{1}{2}g^2(|\phi|^2\phi + \chi^2\phi)=0\cr
&&4m^2\chi + 6mg|\phi|^2 + 2g^2|\phi|^2\chi=0\cr
&&a_\mu|\phi|^2=0\;,
\eea
which have as the only nontrivial solutions (excluding $\phi=\chi=0$)
\be
a_\mu=0,\;\;\;
-\chi=|\phi|=\frac{m}{2g}(3\pm 1).
\ee

We can also obtain purely scalar solutions if we impose $a_\mu=0$, $\phi=\chi$ real, in which case the equations of motion consistently truncate to
\be
-\d^2 \chi+4m^2\chi+6mg\chi^2+2g^2\chi^3=0.\label{onlychi}
\ee

{\bf Vortex ansatz}

We can ask whether there exist vortex solutions. Since there is no possibility for symmetry breaking, this seems unlikely, but we can write an 
ansatz. 

In 3+1 dimensions, vortices are string-like objects, but one can still consider particle-like objects by taking a configuration 
with all the fields constant in one spatial direction, i.e., we can consistently truncate the equations of motion by putting $a_3=\d_3=0$, 
namely looking for vortices that are straight lines in the third direction. We also only consider static solutions $\d_0=0$, and we 
choose the gauge $a_0=0$, which means that we reduce the system of equations to two spatial dimensions as for the usual Abrikosov-Nielsen-Olesen vortex.

The natural ansatz for vortex solutions in polar coordinates is
\be
\phi=f(r)e^{i\theta(\varphi)};\;\;\;
\chi=\chi(r);\;\;\;a_r=a_r(\varphi);\;\;\;a_{\varphi}=a_{\varphi}(r).
\ee
Here $r$ and $\varphi$ are polar coordinates in the 2 dimensional complex plane 1,2 and $\theta $ is the phase of the complex scalar field $\phi$.
Moreover, for an $N$-vortex solution, we have $\theta=N\vp$, where $N=\pm 1,\pm 2, \pm 3,...$ is a winding number.

In order to obtain a finite solution, we need to impose boundary conditions at infinity,
\bea
&& \lim_{r\rightarrow \infty}|\phi|(r)=-\lim_{r\rightarrow\infty}\chi =\frac{m}{2g}(3\pm 1)\cr
&&\lim_{r\rightarrow\infty}D_r\phi(r,\varphi)=0\;\;\;\;\lim_{r\rightarrow\infty}D_\varphi\phi(r,\varphi)=0\cr
&&\lim_{r\rightarrow\infty} \chi'(r)=0\cr
&&\lim_{r\rightarrow\infty}F_{r\varphi}(r,\varphi)=0\;,
\eea
where $F_{r\varphi}$ is the field strength. At $r\rightarrow 0$ we need as usual $|\phi|(r\rightarrow 0)= 0$ in order for $\phi$ to be well-defined, more precisely one finds for the $N$-vortex that
\be
|\phi|(r)\sim r^N\;,
\ee
and now we must impose also $\chi(r)\sim r^\a$ with $\a\geq 0$ for finiteness. From $D_\varphi \phi\rightarrow 0$ at infinity we find
\be
(iN-ia_{\vp})=0\quad \text{i.e}\;\;\lim_{r\to \infty}a_{\vp}(r)=N\;,
\ee
which  implies that the magnetic flux is quantized as usual,
\be
\int\frac{1}{r}F_{r,\vp}rdrd\vp=\int_0^{2\pi}a_{\vp}(\infty)d\vp=2\pi N \,.
\ee
On the other hand,
\be
 D_{r}\phi=0\rightarrow (\dot{\pb}-ia_{r})=0\quad\text{i.e}\quad\lim_{r\to \infty}a_{r}(r)=0\;,
\ee
and finally 
\be
F_{r,\vp}(r)=0\rightarrow \d_r a_{\vp}=0\quad\text{i.e}\quad\lim_{r\to\infty}\d_r a_{\vp}(r)=0\,,
\ee
We see that the above relations imply $a_{\vp}(r\rightarrow\infty)=N$.

Solving the equations of motion for this ansatz is very difficult. We have been unable to find solutions, or to show whether they exist.

\Comment{On the ansatz, the constraint (\ref{constraint}) reduces to 
\be
e^{iN\varphi}\left[\chi\d_\mu f -f\d_\mu \chi+i\chi f(N\d_\mu\varphi-ga_\mu)\right]=0
\ee
which splits into
\bea
&&\chi\d_\mu f=f\d_\mu \chi\cr
&&\chi f(N\d_\mu \varphi-ga_\mu)=0
\eea
Assuming $f\neq 0$, we have a solution $\chi=0$, which however does not satisfy the $\chi$ equation of motion, or otherwise we have
\bea
&& a_\mu=\frac{N}{g}\d_\mu\varphi\cr
&&\chi \d_\mu f =f\d_\mu \chi\Rightarrow \chi f \d_\mu\ln\frac{f}{\chi}=0\Rightarrow f=C\chi
\eea
Substituting in (\ref{chieq}), for consistency with (\ref{phieq}) we must have $C=\pm 1$, i.e. $f=\pm \chi$. 

Therefore, at least on the vortex ansatz, the only possibility is $\chi=|\phi|$, where $\chi$ must satisfy (\ref{onlychi}). We can check that this 
satisfies all the equations of motions, as well as the constraint. 
The only problem is that there is a 
singular field strength, since $a_\mu=(N/g)\d_\mu\varphi$ implies, from $\d_{[\mu}\d_{\nu]}\varphi =2\pi\epsilon_{\mu\nu}\delta^2(r)$, that
\be
F_{\mu\nu}=\frac{2N}{g}\epsilon_{\mu\nu}\delta^2(r).
\ee

In conclusion, we have vortex solutions, but they have a delta function field strength, so are not very physical, though perhaps the can become 
if we regularize (for instance by excising) the region close to $r=0$. }

\subsection{Two-mass deformation}

We can consider a supersymmetric mass deformation of ${\cal N}=4$ SYM that depends on two mass parameters instead of one, with 
superpotential
\be
W=-im\Tr[\Phi_1\Phi^1+\Phi_2\Phi^2]-i\tilde m\Tr[\Phi_3\Phi^3]+\Tr[\Phi_1[\Phi_2,\Phi_3]]\;,
\ee
leading to the bosonic Lagrangean
\bea
{\cal L}_{Bos} &=& \frac{1}{2g^2}{\rm Tr} \left\{\frac{1}{2}
F_{\mu\nu} F^{\mu\nu} + \sum_{m=1}^3\left( { D}_\mu \Phi^{m} \right)
\left( {D}^\mu \overline{\Phi}_{m} \right)+\sum_{m=1}^3\left|\frac{[\Phi_m,\overline{\Phi_m}]}{2}\right|^2
\right.\nn\\
&&\left. + |[\Phi_2,\Phi_3]-2im\Phi_1|^2+|[\Phi_3,\Phi_1]-2im\Phi_2|^2+|[\Phi_1,\Phi_2]-2i\tilde{m}\Phi_3|^2\frac{}{}\right\}.\nn\\
\eea
Again putting $X_{m+3}=0$, so $\Phi_m=X_m$, and writing it terms of $X^\pm$ and $X_3$ and rescaling the fields by $g$, the bosonic Lagrangean becomes
\bea
{\cal L}_{Bos} &=& {\rm Tr} \Bigg\{ \frac{1}{4}
F_{\mu\nu} F^{\mu\nu} + \frac{1}{2}\left( { D}_\mu X_{3} \right) \left( {D}^\mu X_{3} \right) + \frac{1}{2}\left( { D}_\mu X^{+} \right) \left( {D}^\mu X^{-} \right)\nn\\
&& + 2m^2X^{+}X^{-} + 2\tilde{m}^2X_{3}^2  + g(2m+\tilde{m})X_3[X^+,X^-]\nn\\
&& - \frac{g^2}{2}\left(-\frac{1}{4}[X^+,X^-]^2 + [X^+,X_3][X^-,X_3]\right)\Bigg\}.
\eea
Under the same abelianization ansatz as in the one-mass case, we obtain the abelian action
\bea
S&=&N\int d^4x \bigg[+\frac{1}{4}
F_{\mu\nu} F^{\mu\nu} + \left( {D}_\mu \phi \right) \left(\overline{{D}^\mu\phi}\right) + \frac{1}{2}(\partial_{\mu}\chi)^2 + 2\tilde{m}^2\chi^2 + 4m^2|\phi|^2\nn\\
&&\hspace{1cm} + 2(2m + \tilde{m})g|\phi|^2\chi + \frac{1}{2}g^2(\vert\phi\vert^4 + 2\vert\phi\vert^2\chi^2)\bigg]. \label{master}
\eea

The analysis of the consistency of the truncation is exactly the same. From the equations of motion of the scalars $X_m$, we obtain the equations of 
motion of $\chi$ and $\phi$, which are now
\bea
-\d^2\chi + 4\tilde{m}^2\chi + 2g(2m+\tilde{m})|\phi|^2 + 2g^2|\phi|^2\chi &=& 0,\nn\\
-D^2\phi + 4m^2\phi + 2g(2m+\tilde{m})\chi\phi + g^2(|\phi|^2\phi + \chi^2\phi) &=& 0.
\label{masteqs}
\eea
From the equation of motion of the gauge fields, we obtain for $a=3$ the equation of motion of the reduced gauge field, which is the same (\ref{gaugeeq}).

If $2m+\tilde m\neq 0$, we cannot have the same vacuum solutions, or solutions with $-\chi=|\phi|$ anymore, since then the $\chi$ and $\phi$ 
equations of motion are incompatible due to the mass term. It is also again not consistent to put $\chi$ to a constant while $\phi$ is general, 
since there is still a term linear in $\chi$ in the action. But with the more general ansatz $\chi=a|\phi|$, we obtain for the vacuum solutions
(multiply the first equation in (\ref{masteqs}) by $a$ and subtract them)
\bea
a&=&-\sqrt{\frac{4m^2+g^2|\phi|^2}{4\tilde m^2+g^2|\phi|^2}}\cr
g(2m+\tilde m)|\phi|&=&-a(2\tilde m^2+g^2|\phi|^2).\label{newvacsol}
\eea

For vortices, the same ansatz as in the one-mass case applies, the only difference is that at infinity, $|\phi|$ and $\chi$ need to go to the new 
vacuum solution, defined in (\ref{newvacsol}). We have again been unable to find solutions or to prove whether they exist.

However, now a new possibility appears if $2m+\tilde m=0$. The resulting action, 
\bea
S&=&N\int d^4x \bigg[+\frac{1}{4}
F_{\mu\nu} F^{\mu\nu} + \left( {D}_\mu \phi \right) \left(\overline{{D}^\mu\phi}\right) + \frac{1}{2}(\partial_{\mu}\chi)^2 + 8m^2\chi^2 + 4m^2|\phi|^2\nn\\
&&\hspace{1cm} + \frac{1}{2}g^2(\vert\phi\vert^4 + 2\vert\phi\vert^2\chi^2)\bigg]\;,
\eea
has no cubic term, so now the equation of motion for $\chi$,
\bea
-\d^2\chi + 16m^2\chi + 2g^2|\phi|^2\chi &=& 0,
\eea
admits the consistent truncation $\chi=0$, after which we obtain simply massive scalar QED,
\bea
S=N\int d^4x \bigg[+\frac{1}{4}
F_{\mu\nu} F^{\mu\nu} + \left( {D}_\mu \phi \right) \left(\overline{{D}^\mu\phi}\right) + 4m^2|\phi|^2 + \frac{1}{2}g^2\vert\phi\vert^4\bigg].
\eea

\section{FI deformation of ${\cal N}=4$ SYM and abelian reduction to Landau-Ginzburg.}

In order to obtain a Higgs potential in a supersymmetric gauge theory, one usually considers a Fayet-Iliopoulos (FI) term for an abelian theory. 
Therefore in this section we consider the two-mass deformation of section 2.2 (though it will not matter, since the $\tilde m$ will drop out of our 
calculation anyway), and on top of it, an FI deformation.

For an abelian vector field, with real superfield $V$, in the Wess-Zumino gauge we have
\be
V=-\bar\theta\sigma^\mu \theta A_\mu +i\theta^2(\bar\theta\bar\psi)-i\bar\theta^2(\theta\psi)+\frac{1}{2}\theta^2\bar\theta^2 D\;,
\ee
and the gauge-scalar super-interaction term in the Lagrangean is
\be
\int d^2\theta d^2\bar\theta \Phi^\dagger\Phi V\;,
\ee
and the FI term is 
\be
-\xi\int d^2\theta d^2\bar\theta V.
\ee
Solving for the auxiliary scalar $D$, one gets
\be
D=-\xi+\phi^\dagger\phi\;,
\ee
where $\phi$ is the first component of the superfield $\Phi$. The scalar potential is $D^2$. Then $\xi<0$ gives spontaneous supersymmetry breaking, while 
$\xi>0$ gives spontaneous gauge symmetry breaking. 

In ${\cal N}=4$ SYM, we have 3 chiral superfields $\Phi_1,\Phi_2,\Phi_3$ and a real (vector) superfield $V$, all in the adjoint of the gauge group $SU(N)$. 
We can deform the (already massively deformed) 
theory by adding a FI term in the same $U(1)$ direction as the gauge field for the abelian reduction, i.e. $V=vJ_3$, 
since $A_\mu=a_\mu J_3$. But then we also want the complex scalar to be gauged with 
respect to the same direction, so as to have $D=-\xi+\phi^\dagger\phi$. This is not possible in the reduction ansatz of the previous section, 
where the D-terms vanish on the truncation, since $\Phi_i$ are real, so $[\Phi_i,\Phi_i^\dagger]=0$. 

Therefore we need to consider instead the abelian reduction ansatz
\be
\Phi_1=\phi J^+,\;\;\;
\Phi_1^\dagger=\phi^* J^-;\;\;\;
\Phi_2=\chi J_3;\;\;\;
\Phi_3=0;\;\;\; A_\mu=a_\mu J_3.\label{modansatz}
\ee

The FI term can be written in an $SU(N)$ invariant way as
\be
\int d^2\theta d^2\bar\theta \Tr[\Xi V]\;,
\ee
where $\Xi$ is now a constant matrix, taken on the abelian reduction ansatz to be $\Xi=\xi J_3$,
and the scalar-gauge supersymmetric coupling is  
\be
\int d^2\theta d^2\bar\theta \sum_i \Tr[\Phi_i^\dagger e^{-gV}\Phi_i].
\ee
In total, the bosonic Lagrangean of the deformed theory is now
\bea
{\cal L}_{Bos} 
&=& \frac{1}{2}{\rm Tr} \bigg\{\frac{1}{2}
F_{\mu\nu} F^{\mu\nu} + \left({D}_\mu\Phi^{m}\right)\left({D}^\mu\overline{\Phi}_{m}\right) \cr
&&+\left|\frac{g[\Phi_1,\overline{\Phi_1}]}{2}-\Xi\right|^2
+\left|\frac{g[\Phi_2,\overline{\Phi_2}]}{2}\right|^2+\left|\frac{g[\Phi_3,\overline{\Phi_3}]}{2}\right|^2\nn\\
 &&+ |g[\Phi_2,\Phi_3]-2im\Phi_1|^2+|g[\Phi_3,\Phi_1]-2im\Phi_2|^2+|g[\Phi_1,\Phi_2]-2i\tilde{m}\Phi_3|^2\bigg\}.
\eea

The D-term equation is now, on the reduction ansatz ($[\Phi_2^\dagger,\Phi_2]=0$)
\be
-D=\xi-g\phi^\dagger\phi.
\ee

The covariant derivative reduces as before to
\be
D_\mu \Phi_1=(\d_\mu\phi-ig a_\mu\phi)J^+;\;\;\;
D_\mu \Phi_2=\d_\mu\chi J_3\;,
\ee
the kinetic term to 
\be
N\left[\frac{1}{4}F_{\mu\nu}^2+|D_\mu\phi|^2+\frac{1}{2}(\d_\mu\chi)^2\right]\;,
\ee
and the potential reduces to \footnote{$\Tr[J^+J^-]=2N$, $\Tr[J_3 J_3]=N$, $\Tr[J^+J^+]=\Tr[J^-J^-]=\Tr[J^+J^3]=0$}
\be
V=\frac{N}{2}\left[4m^2(2|\phi|^2+\chi^2)+(g\phi^*\phi-\xi)^2+2g^2\chi^2|\phi|^2\right]\;,
\ee
coming from the mass term in the superpotential, D-term, and commutator term in the superpotential respectively. 
We see that indeed, if $\xi>0$, we have a negative mass squared contribution to the potential for $\phi$, and moreover, since we already have a 
mass term, we have a relativistic  Landau-Ginzburg theory, with a parameter $\xi$ that controls whether the mass squared of $\phi$, $M^2$, 
is positive or negative
(like in $V\sim  (g-g_c)\phi^2+\lambda \phi^4$). Indeed, now for $-2\xi g^2+8m^2<0$ we have the symmetry-breaking abelian-Higgs model, 
otherwise we have a massive $\phi^4$ theory.

Note also that now, with the new truncation ansatz, there is no term linear in $\chi $ (term cubic in all the fields), 
since all the cubic terms involve all the fields $\Phi_1,\Phi_2,\Phi_3$, and $\Phi_3=0$ on the ansatz, so the further truncation to $\chi=0$ 
will be consistent.

{\bf Consistency of the truncation}

The consistency of the truncation works in exactly the same way as in the previous section. 
The $\Phi_1^\dagger$ equation of motion for $\Phi_3=0$ is 
\be
-\frac{1}{2}D^2\Phi_1+2m^2\Phi_1+\frac{g^2}{4}[[\Phi_1,\Phi_1^\dagger],\Phi_1]+\frac{g^2}{2}[\Phi_1,\Phi_2],\Phi_2^\dagger]-\frac{g}{2}[\Xi,\Phi_1]=0\;,
\ee
and on the reduction ansatz it reduces to 
\be
J^+\left(-\frac{1}{2}D^2\phi+2m^2\phi-\frac{g}{2}\xi \phi+\frac{g^2}{2}\chi^2\phi+\frac{g^2}{2}|\phi|^2\phi\right)=0\;,
\ee
which is the equation of motion for $\phi$ in the reduced model (times an overall 1/2). 

The equation of motion for $\Phi_2^\dagger$ for $\Phi_3=0$ is 
\be
-\frac{1}{2}D^2\Phi_2+2m^2\Phi_2+\frac{g^2}{4}[[\Phi_2,\Phi_2^\dagger],\Phi_2]+\frac{g^2}{2}[\Phi_2,\Phi_1],\Phi_1^\dagger]=0\;,
\ee
which on the reduction ansatz reduces to
\be
J_3\left(-\frac{1}{2}\d^2\chi+2m^2\chi+g^2|\phi|^2\chi\right)=0\;,
\ee
which is the equation of motion for $\chi$ in the reduced model (times an overall 1/2).
The equation of motion for $\Phi_3^\dagger$ is satisfied for $\Phi_3=0$ on the reduction ansatz. 

As before, the gauge field equation of motion reduces to the abelian gauge field equation of motion for $a=3$, exactly like in 
(\ref{Aeom}). Potential troublesome terms, like there, would be 
\be
\delta^{mn}g\Phi_m^b D_\mu \Phi_m^c {f^a}_{bc}\;,
\ee
in the equation for $a=+$, but for which we need $(bc)=(+3)$, and with our ansatz, that would require $(mn)=(12)$, but this is excluded due to 
the $\delta^{mn}$.

In conclusion, the truncation is again consistent.

Moreover, now $\chi=0$ is a further consistent truncation, that leaves simply the relativistic Landau-Ginzburg model.

\section{Vortex solutions and a theorem}

Given that we have an LG model, we certainly have the usual Abrikosov-Nielsen-Olesen vortices in the abelian-Higgs phase.
The general multivortex solutions are as usual (in complex coordinates $z,\bar z$)
\bea
\phi(z,\bar z)&=&ve^{-\frac{\psi(z,\bar z)}{2}}H_0(z)\cr
a_{\bar z}&=&\frac{i}{2}\bar\d \psi(z,\bar z)\;,
\eea
where $H_0(z)=\prod_{i=1}^n (z-z_i)$ and $\psi$ satisfies the equation 
\be
\d\bar\d\psi=M^2(1-e^{-\psi}|H_0(z)|^2)\;,
\ee
with boundary conditions at $|z|\rightarrow \infty$ requiring $\psi\rightarrow \log |H_0|^2$, where $M$ is the mass of $\phi$ and $v$ its VEV. 
We embed them into ${\cal N}=4$ SYM through the abelianization ansatz (\ref{modansatz}) like in \cite{Mohammed:2012rd,Mohammed:2010eb}.

But an interesting possibility that we want to study is whether we can have vortex solutions with a nontrivial $\chi$.

\subsection{BPS condition and vortices}

We start by analyzing BPS solutions.
As before, we consider static solutions with trivial $x_3$ direction ($\d_3=\d_0=0$) in the axial gauge $a_0=0$, and note that $a_3=0$ is a 
consistent truncation, thus reducing the system to a 2 dimensional one. 

The energy density is then given by
\be
\frac{{\cal E}}{N}=\frac{F_{0i}^2}{2}+\frac{F_{12}^2}{2}+|D_0\phi|^2+|D_i\phi|^2+\frac{1}{2}(\partial_i\chi)^2
+2m^2(2|\phi|^2+\chi^2)+g^2|\phi|^2\chi^2+\frac{1}{2}\left(g|\phi|^2-\xi\right)^2\,,
\ee 
with $i=1,2$. Notice that the energy is greater or equal to zero, since it is a sum of positive terms.

One can rewrite it by completing squares in the usual way as
\bea
\frac{\cal E}{N}&=& \frac{F_{0i}^2}{2}+\frac{1}{2}\left(F_{12}+g|\phi|^2-\xi+\frac{4m^2+g^2\chi^2}{g}\right)^2
-\left(-\xi+\frac{4m^2+g^2\chi^2}{g}\right)^2\nn\\
&&-F_{12}\left(-\xi+\frac{4m^2+g^2\chi^2}{g}\right)+\xi^2+|D_0\phi|^2+|D_+\phi|^2+\frac{1}{2}(\partial_i\chi)^2+2m^2\chi^2\nn\\
&&-i\epsilon^{ij}\partial_i(\phi^{\dagger}D_j\phi).
\eea
When $\chi=0$, the third term becomes a number, the fourth becomes a topological index after integration, 
and the last should vanish after the spatial integration, because of the boundary conditions at infinity (becoming a 
surface term). Hence, as usual, the minimal (BPS) energy in a given topological charge (vortex charge) sector is reached for  
\bea
F_{0i}&=&0\,,\label{f0i}\\
D_+\phi&=&\chi=0\,\label{dplus},\\
F_{12}&=&\xi-g|\phi|^2-\frac{4m^2+g^2\chi^2}{g}\,.\label{f12}
\eea
We see that we can only satisfy these BPS (minimal energy) conditions if $\chi=0$, in which case we obtain the usual Landau-Ginzburg model 
(with abelian-Higgs phase), so the only BPS vortices are the usual Avrikosov-Nielsen-Olesen ones.

\subsection{Non-BPS vortices}

There is still the possibility that there are non-BPS vortex solutions with nontrivial $\chi$. 

The bosonic equations of motion of the abelian LG-like  action are
\bea 
D_{\mu}D^{\mu}\phi&=&4m^2\phi+g^2\phi\left(\chi^2+|\phi|^2-\zeta\right)\,,\nn\\
\pd_{\mu}\pd^{\mu}\chi&=&2\chi(2m^2+g^2|\phi|^2)\,,\nn\\
\pd_ {\mu}\pd^{\mu}a^{\rho}-\pd^{\rho}\pd^{\mu}a_{\mu}&=&ig^2(D^{\rho}\phi\,\phi^*-\phi\,D^{\rho}\phi^*)\,.
\eea
We again consider static solutions, with trivial $x_3$ dependence ($\d_3=0$), in the axial gauge $a_0=0$, and in the case of the consistent 
truncation $a_3=0$. Denoting by $(r,\a)$ the cylindrical coordinates parametrizing the plane  $(x_1,x_2)$, the 
vortex ansatz is
\be
\phi=|\phi|(r) e^{i\theta},\;\;\; \theta= N\a,\;\;\;
\chi=\chi(r).
\ee
We take $N=1$, for the one-vortex solution, and examine the asymptotics of possible vortex solutions.

At $r\rightarrow\infty$, since the field $\chi$ is massive, as we can easily see from the equations of motion, for it we have asymptotically
\be
\chi\sim A e^{-Mr}\;,
\ee
where $M$ is the mass of the field, $M^2=2(2m^2+g^2v^2)$, and $v$ is the VEV of $\phi$. Since we have an exponentially small $\chi$, it is 
guaranteed to introduce a finite contribution to the energy, even though it is not BPS, so a priori one could have expected an infinite contribution.

For $\phi$, since we have an exponentially small $\chi$, we can take the usual BPS Abrikosov-Nielsen-Olesen vortex solution, meaning that we can write
at $r\rightarrow \infty$
\be
|\phi|-v\sim \frac{A_1}{r^n}
\ee
and the gauge field $a_i$ as in the BPS solution. 

At $r\rightarrow 0$, we would like to have $|\phi|\sim K r^N$, i.e $|\phi|\sim r$ for $N=1$ vortex, as for the Abrikosov-Nielsen-Olesen vortex, and
we check whether this is possible (consistent with the equations of motion). 

An exact solution to the free massive equation for a real scalar in 2+1 dimension is $\phi=A K_0(Mr)$, where $K_\nu$ are Bessel functions,
which goes to $A\sqrt{\pi/(2Mr)} e^{-Mr}$ at $r\rightarrow\infty$ and to $-A\ln (Mr/2)$ at $r\rightarrow 0$, but such a solution gives a divergent 
energy at $r=0$, since the contribution to the energy from the $r=0$ endpoint for integration is
\be
\Delta E\sim \int_0 (\d_i\chi)^2 (2\pi r dr)\sim \int_0 (\chi')^2 (2\pi r dr)\sim 2\pi \int_0 dr/r\rightarrow\infty.
\ee

But we note that near $r=0$ we can instead have a solution with a well-defined Taylor expansion, with 
\be
\chi\simeq A+Br +Cr^2+...
\ee
Plugging it into the equation of motion for $\chi$ 
\be
\nabla\chi=\chi(\tilde M^2+\tilde g^2 |\phi|^2)\;,
\ee
with $\tilde g =g\sqrt{2}$ and $\tilde M=2M$, we find
\be
2C+\frac{B}{r}+2C+...=(A+Br+Cr^2...)(\tilde M^2+\tilde g^2 K^2 r^{2N}+...)\;,
\ee
meaning that $B=0$ and $C=A\tilde M^2/4=AM^2$, so 
\be
\chi\sim A\left(1+M^2 r^2+...\right).
\ee
The equation of motion for $\phi$ is of the type
\be
D^\mu D_\mu \phi=\phi(usual)+\phi g^2\chi^2.
\ee
Since $\chi$ is proportional to the arbitrary constant $A$, that can be made as small as we like, we can treat the field $\chi$ near $r=0$ as a 
small perturbation that just redefines a bit the negative mass squared of $\phi$ in the trivial vacuum. 
This solution near $r=0$ also has finite energy. 

However, the problem is that we cannot have a solution that has finite energy at both $r\rightarrow 0$ and $r\rightarrow \infty$, 
since the equation of motion for $\chi$ implies 
\be
\frac{\chi''(r)}{\chi}+\frac{1}{r}\frac{\chi'}{\chi} >0\;
\ee
so we could only have the solution $e^{-Mr}$ at infinity if it goes over to $\ln(Mr/2)$ (which is decreasing with $r$), and the solution  with $A+Cr^2$ at 
$r=0$ goes over to $e^{+Mr}$ at infinity. If there would be a solution starting as $A+Cr^2$ and ending as $e^{-Mr}$, it would need to have a maximum, 
i.e. $\chi'=0$ in between, which would require that $\chi''<0$ and $\chi'=0$ somewhere. 
Thus there are no vortex solutions with $\chi$ nontrivial and finite energy.

In conclusion, we have proved that there are no vortex solutions with nontrivial $\chi$, either BPS or non-BPS, which is the theorem 
alluded to in the title of the section.

\section{Conclusions}

In this paper we have studied possible abelian reductions of 3+1 dimensional 
deformed ${\cal N}=4$ SYM to the relativistic Landau-Ginzburg model, motivated by 
the similar result for the 2+1 dimensional ABJM model, and by possible applications to condensed matter via AdS/CMT.

We have found that taking a mass deformation with one or two mass parameters, we can obtain just a scalar QED coupled to a real scalar $\chi$, that 
cannot be consistently truncated to a constant except in a special case. We have studied possible solutions to these models, 
but no new vortex solutions were found.

By taking instead a FI term deformation of the theory deformed with a single mass (we can take a two-mass deformation, but the second mass 
drops out when we take the reduction ansatz), we can reduce to the relativistic Landau-Ginzburg model
coupled to a real scalar field $\chi$, and the truncation is consistent, and moreover $\chi=0$ is also a consistent truncation. 
We have proven that in the resulting theory there are no vortices with nontrivial $\chi$ scalar profile.

We have reduced deformed ${\cal N}=4$ SYM to a relativistic LG theory having in mind applications to AdS/CMT, as was done in 
2+1 dimensions in \cite{Mohammed:2012gi,Mohammed:2012rd} for the ABJM model. The LG theory appears as an effective field theory in 
condensed matter systems, so it would be nice see whether we can mimic the reduction in degrees of freedom that leads to LG in a condensed matter
system, from the point of view of ${\cal N}=4$ SYM, viewed as a toy model  for it. We also note that 
vortex solutions play an important role in the CMT description, 
in particular for the description of physics near a quantum phase transition, see e.g. \cite{Sachdev:2011wg}. We leave the study of condensed matter 
implications of the abelianization and vortex solutions for further work.

Towards that goal, one needs to understand the effect of the abelian truncation on the gravity dual. This will help make concrete the duality to 
condensed matter systems, with the final goal of using the gravity dual for the LG theory, and understanding the role of ${\cal N}=4$ SYM for 
the AdS/CMT correspondence. Note that, like in the case of the ABJM theory analyzed in \cite{Mohammed:2012gi,Mohammed:2012rd}, the truncation 
considered here does not involve simply an abelian ($U(1)$) version of the SYM, but rather a subsector of nonabelian matrices defined by the representations
$J_i$ of the fuzzy 2-sphere ($SU(2)$) algebra, equivalent \cite{Nastase:2009zu} to the representations of another fuzzy 2-sphere algebra in terms 
of the matrices $G_\a$ used in \cite{Mohammed:2012gi,Mohammed:2012rd}. The intrinsically nonabelian nature of the matrices used in the reduction, 
with ${\cal O}(N)$ nontrivial elements turned on at large $N$, means it is likely that we can use a nontrivial restriction of the gravity dual, as it was 
argued in the ABJM case. 

It would also be interesting to consider an abelian truncation to a supersymmetric model, i.e. an extension of the truncation that includes fermions and 
preserves some of the supersymmetry, like it was done in \cite{Murugan:2013jm} for the ABJM case.
Note that vortex solutions of models with FI terms were related to superconductivity in Seiberg-Witten theory \cite{Vainshtein:2000hu}.

{\bf Acknowledgements}

We would like to thank Nathan Berkovits and Andrei Mikhailov for discussions.
The research of HN is supported in part by CNPQ grant 301709/2013-0
and FAPESP grant 2013/14152-7. CLA thanks Jonathan Shock for helping by searching for numerical solutions. The research of CLA was 
supported in part by CAPES PhD scholarship and now by NRF CPRR grantholder posdoctoral fellowship. The research of CC 
is supported in part by CNPQ grant 501043/2012-8.

\bibliography{SYMred}
\bibliographystyle{utphys}

\end{document}